\documentclass{iaus}
\usepackage{graphicx}

\newcommand{\topp}{{\em top}}
\newcommand{\bott}{{\em bottom}}
\newcommand{\lee}{{\em left}}

\newcommand{\panel}{{\em panel}}
\newcommand{\panels}{{\em panels}}

\newcommand{\psicen}{$\psi$~Cen}
\newcommand{\arcas}{AR~Cas}
\newcommand{\betaaur}{$\beta$~Aur}
\newcommand{\psicenn}{$\psi$~Centauri}
\newcommand{\arcass}{AR~Cassiopeiae}
\newcommand{\betaaurr}{$\beta$~Aurigae}

\newcommand{\sint}{$\sigma_{\rm int}$}

\newcommand{\wire}{{\sc wire}}
\newcommand{\smei}{{\sc smei}}
\newcommand{\coriolis}{{\sc coriolis}}

\newcommand{\kepler}{{\sc kepler}}
\newcommand{\corot}{{\sc corot}}

\title[Eclipsing Binary Stars from Space] 
{Eclipsing Binary Stars from Space}

\author[Bruntt \& Southworth]   
{Hans Bruntt$^1$ and John Southworth$^2$}

\affiliation{$^1$School of Physics, University of Sydney, Australia
\break email: hans@bruntt.dk\\[\affilskip]
$^2$Department of Physics, University of Warwick, UK\break email: jkt@astro.keele.ac.uk}

\pubyear{2007}
\volume{240}  
\pagerange{2500-2503}
\date{?? and in revised form ??}
\jname{Binary Stars as Critical Tools \& Tests in Contemporary Astrophysics}
\editors{W.\ Hartkopf, E.\ Guinan \& P. Harmanec, eds.}
\begin{document}

\maketitle

\begin{abstract}
We have begun a programme to obtain 
high-precision photometry of bright detached eclipsing binary (dEB) stars
with the Wide field InfraRed Explorer (\wire) satellite \cite[(Bruntt \& Buzasi 2006)]{bruntt06a}.
Due to the small aperture of \wire\ only stars brighter than $V=6$ can be observed.
We are collecting data for about a dozen dEB targets and here we present preliminary results for three of them.
We have chosen dEBs with primary components of B and early A type. 
One of our aims is to combine the information from the light curve analyses of 
the eclipses with asteroseismic information from the analysis of 
the pulsation of the primary component. 
\end{abstract}

\firstsection 
\section{Introduction}

The study of detached eclipsing binaries is of fundamental importance to stellar
astronomy as a way of measuring accurately the parameters of normal stars from basic observational data
\cite[(Andersen 1991)]{andersen91}. The masses and radii of         
detached eclipsing binary (dEB) stars can
be measured to accuracies better than 1\%,      
and the effective temperatures and luminosities can be obtained from spectral analysis or the use of
photometric calibrations. An important use of these data is in the calibration of theoretical models
of stellar evolution,
particularly if the component stars are of quite different mass or evolutionary stage
\cite[(\eg\ Andersen et al. 1991)]{andersenetal1991}.


The importance of the physical effects included in theoretical models like the extent
of convective core overshooting and the efficiency of convective energy transport (mixing length), 
need to be constrained by observations to increase the predictive power of the models. 
This may be done by measuring increasingly accurate absolute properties of dEBs, 
which requires much improved observational data. 
It is arguably more difficult to significantly improve the quality and quantity of light curves,
rather than radial velocity curves, because of the large amount of telescope time needed for each system.
The \kepler\ \cite[(Basri et al.\ 2005)]{kepler05}                 
and \corot\  \cite[(Baglin et al.\ 2001)]{corot01} space missions  
will help to solve this problem in the future by obtaining accurate and extensive light 
curves of a significant number of dEBs. 
The results presented here from the \wire\ satellite give an 
indication of the level of accuracy that future space missions will provide.

  \begin{figure}
   \includegraphics[width=13.5cm]{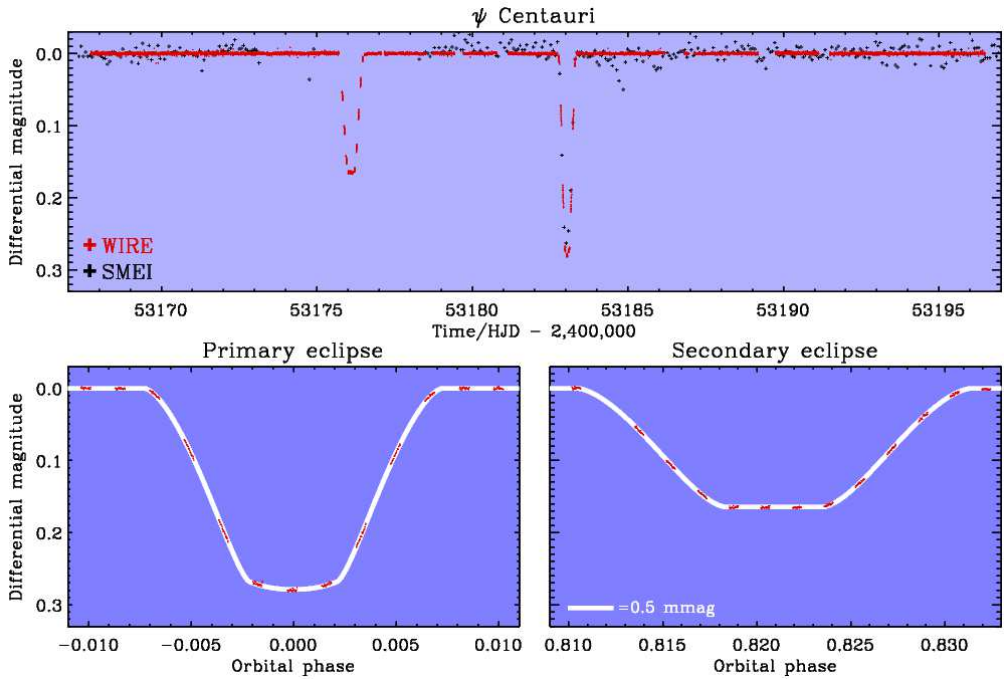}
    \caption{The light curve of \psicen. In the \topp\ \panel\ red points are from
\wire\ and black points are from \smei. The \bott\ \panels\ show the phased
light curve around the primary and secondary eclipse.}
\label{fig:psicen}
  \end{figure}

\section{Three dEBs observed with \wire}

We will present preliminary results for three dEBs observed with \wire:
\psicenn\ is a newly discovered system and has a long 
orbital period, while \arcass\ and \betaaurr\ are known systems with shallow eclipses. 
For these reasons observations with ground-based telescopes are difficult.
\psicen\ and \arcas\ are particularly interesting systems since they 
have secondary components which have much smaller masses and radii 
than the primary stars so the eclipses are total. 
This allows us to determine the radii with increased accuracy compared to 
\betaaur\ which has partial eclipses.
We are collecting new spectroscopic data for all systems, 
and so we will be able to provide strict constraints on the predictions of theoretical models.
The basic properties of the dEB systems we present here are given 
in Table~\ref{tab:bruntt240a} below and their light curves are shown in {Figs.~1--3}.

\begin{table}[b]\def~{\hphantom{0}}
  \caption{Properties of the targets: Spectral type, magnitude,
and period, time observed with \wire, and precision per data point.}
  \label{tab:bruntt240a}
  \begin{center}
  \begin{tabular}{l|cccc}  \hline
                     & \psicen  & \arcas  & \betaaur \\ \hline
Spectral type        &  A0\,IV  &  B4\,IV &  A2\,IV  \\ 
$V$                  &   4.0    &  4.9    &  1.9     \\
Period [d]           &  38.813  &  6.066  &  3.960   \\ 
$\Delta T_{\rm obs}$ &  28.7    &  26.4   &  20.9    \\
\sint\ [mmag]        &   1.0    &  1.1    &  0.1     \\
  \end{tabular}
 \end{center}
\end{table}


\subsection{A new dEB: \psicenn}\label{sec:psicen}
Serendipity led to the discovery that 
\psicen\ is a dEB as it was observed as a secondary target for another observing program. 
We used photometry with a time baseline of two years 
from the Solar Mass
Ejection Imager \cite[(\smei; Webb et al.\ 2006)]{webb06}
on the Coriolis satellite to determine the period of 38.813\,d. 

The light curve of \psicen\ from \wire\ is shown with red points 
in the \topp\ \panel\ in Fig.~\ref{fig:psicen}. 
The \smei\ data that was collected at the same time as the discovery light curve from 
\wire\ are shown with black points.                      
The details of the primary and secondary eclipses are shown in the two \bott\ \panels\ and
the white curve is our best-fitting model light curve.   
\cite[Bruntt et al.\ (2006)]{bruntt06b} have made a detailed analysis of the light curve.
From Monte Carlo simulations we find that the fractional radii of the two stars are
determined to unprecedented accuracies of 0.1\% and 0.2\% (random errors). 
\psicen\ is an interesting system since its brightest component is located in the region of the
Hertzsprung-Russell diagram between the blue edge of the instability strip and the
region of $g$-mode oscillations seen in slowly pulsating B stars
\cite[(\eg\ Pamyathnykh 1999)]{pamyat1999}. 
In the Fourier spectrum we detect two low-frequency
modes that we interpret as global oscillation $g$-modes in the primary star.

  \begin{figure}
   \includegraphics[width=13.5cm]{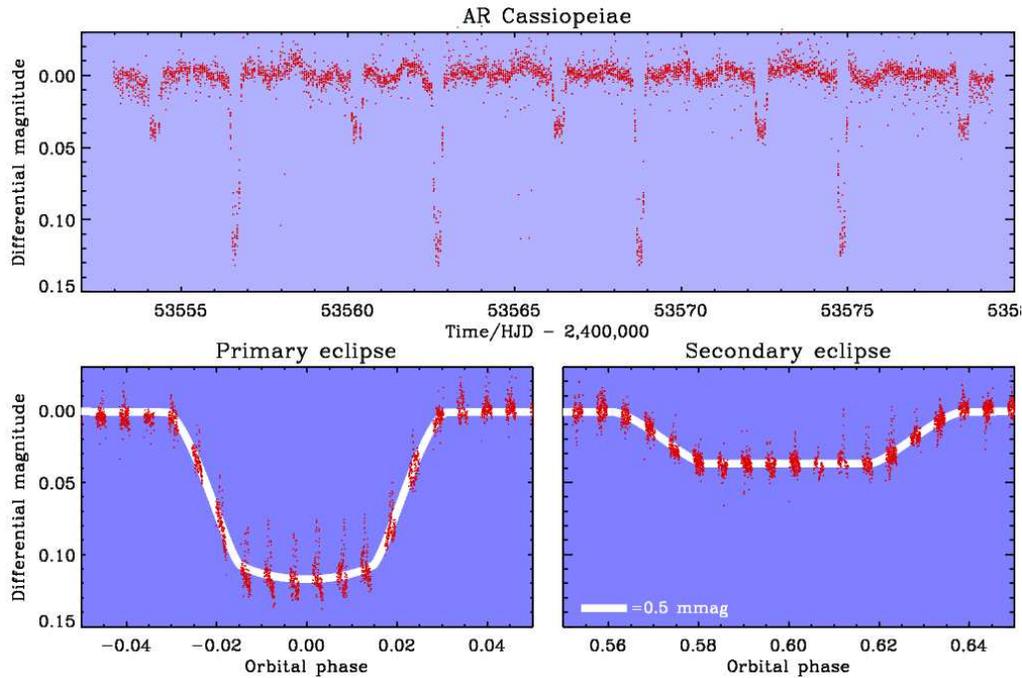}
    \caption{The light curve of \arcas\ is shown 
in the \topp\ \panel\ (every fifth data point is plotted).
The \bott\ \panels\ show the phased light curve around the primary and secondary eclipses.
This is a preliminary data reduction and systematic offsets are seen in the light curve.}
\label{fig:arcas}
  \end{figure}

\subsection{\arcass}\label{sec:arcas}
The light curve of \arcas\ from \wire\ is shown 
in the \topp\ \panel\ in Fig.~\ref{fig:arcas} and the details of the
primary and secondary eclipses are shown in the \bott\ \panels. This 
is a preliminary result and systematic instrumental effects 
have not yet been fully removed. However, intrinsic variations
in brightness are clearly present. Fourier analysis of the time series
after subtracting the best-fitting light curve solution shows the presence
of several pulsation modes with periods from 0.5 to 2.0 days. In addition, there is a brightness modulation
at the rotational period of the primary component, which could be due to surface inhomogeneities in this
star. The spectral type of \arcas\ is B4\,IV + A6\,V and the brightness variations 
are possibly due to SPB-type pulsation in the primary star.

\subsection{\betaaurr}\label{sec:betaaur}
\begin{figure}
   \includegraphics[width=13.5cm]{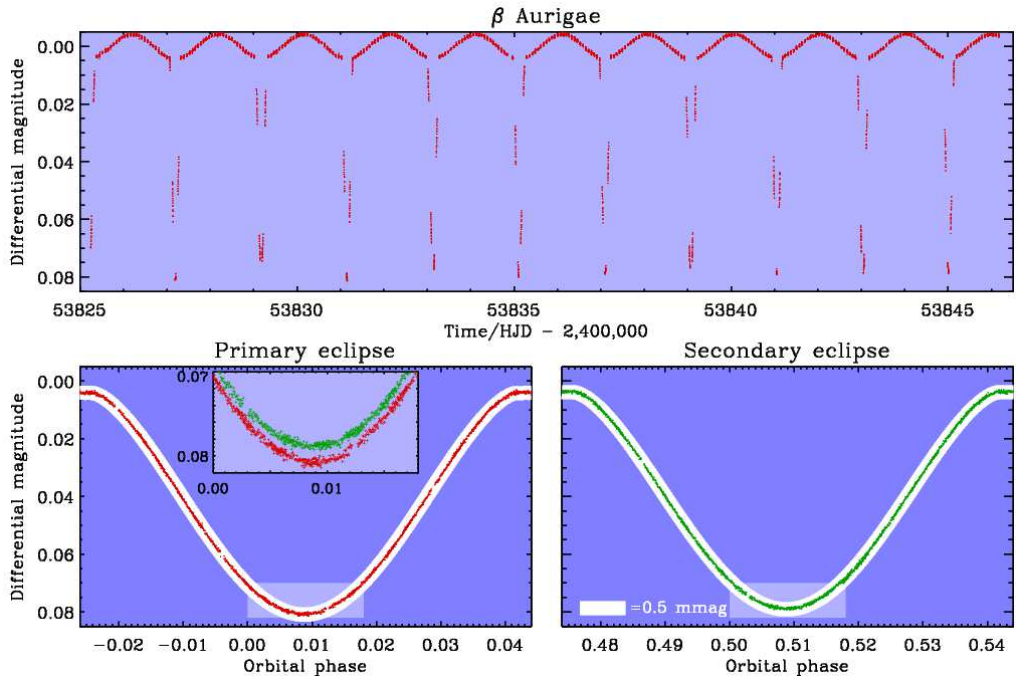}
    \caption{The light curve of \betaaur\ from \wire\ is shown 
in the \topp\ \panel\ (every fifth data point is plotted).
The \bott\ \panels\ show the phased light curve around the primary and secondary eclipses.
The slight difference in the depths of the primary and secondary eclipses is seen
in the the inset in the \bott\ \lee\ \panel.}
\label{fig:betaaur}
  \end{figure}

The light curve from \wire\ of \betaaur\ is shown in Fig.~\ref{fig:betaaur}.
The data have only been obtained very recently, and shows 11 eclipses over a time span of 
21 days. The phased light curve and a preliminary light curve model
are shown in the \bott\ \panels in Fig.~\ref{fig:betaaur}.
The inset in the \bott\ \lee\ \panel\ shows the slightly 
different depths of the primary (red points) and secondary
(green points) eclipses.


 \begin{acknowledgments}
We thank Derek L.\ Buzasi (and the \wire\ team), 
Jens~Viggo Clausen, Guillermo Torres, Alan Penny (and the \coriolis\ \& \smei\ teams), 
Terry Moon, John Innis, Donald W.\ Kurtz, Thebe Medupe, Francois van Wyk, and
Gerald Handler for providing valuable input and making this project possible.
 \end{acknowledgments}

\end{document}